\documentclass[twocolumn]{jpsj3}
\usepackage{txfonts}
\usepackage{braket}

\title{Theory of Self-Induced Vortex State in Ferromagnetic Superconductors}

\author{Hiroaki \textsc{Kusunose}\thanks{hk@ehime-u.ac.jp} and Yuji \textsc{Kimoto}
}

\inst{Department of Physics, Ehime University, Matsuyama, Ehime 790-8577, Japan
}

\abst{
We have developed the quasi-classical formalism for a self-induced vortex state in ferromagnetic superconductors.
By combining the spatially averaged approximation of the superconductivity with the phenomenological form of the ferromagnetism, we demonstrate the thermodynamic properties of the $p$-wave triplet equal-spin-pairing state with the spontaneous vortex lattice.
The occurrence condition for the self-induced vortex state is discussed within this formalism.
The comparison of the calculated quantities with the observed ones indicates that the self-induced vortex state indeed occurs in UCoGe, and its pairing symmetry seems to be the A$_2$-type with a small spin difference in the gap magnitudes.
In the vicinity of the ferromagnetic quantum critical point with a large uniform susceptibility, it is possible to exhibit the type-II to the type-I transition in the magnetization process.
}

\kword{ferromagnetic superconductor, self-induced vortex state, Pesch approximation}

\begin{document}
\maketitle

\section{Introduction}

Interplay between ferromagnetism (FM) and superconductivity (SC) has long attracted attention, and many candidates of their coexistence have been addressed so far\cite{Matthias58,Fertig77,Bauernfeind96,Bulaevskii85,Canfield96,Felner97}.
A paradigm shift has occurred by the discovery of a series of the uranium compounds, such as UGe$_2$\cite{Saxena00}, URhGe\cite{Aoki01}, and UCoGe\cite{Huy07}, which have been investigated extensively in the last decades\cite{Aoki12}.
The coexistence of the FM and the SC has also been proposed in the two-dimensional interface between two bulk insulators, LaAlO$_3$ and SrTiO$_3$\cite{Dikin11}.

These uranium compounds share the zig-zag structure, and the strong Ising anisotropy, leading to the Ising-like FM orders.
UGe$_2$ has much higher FM transition temperature $T_{\rm FM}$ than SC one $T_c$, while they are comparable in UCoGe.
The relation of $T_{\rm FM}$ and $T_c$ in URhGe is intermediate as compared with the above two compounds.
Therefore, these series of compounds provide a good playground for mutual comparison, and UCoGe is the most fascinating material to address the interplay between the FM and the SC.

The NQR/NMR studies in UCoGe have revealed that the longitudinal critical FM fluctuations play an essential role to stabilize the SC state\cite{Ohta10,Hattori12}.
Moreover, the upper critical field $H_{c2}$ is far beyond the Pauli limit\cite{Aoki12,Aoki13}, and spontaneous magnetic moment accompanies large internal magnetic field.
All of these observations suggest the equal-spin-pairing (ESP) state $\Delta_{\uparrow\uparrow}$, $\Delta_{\downarrow\downarrow}\ne0$ in the superconducting phase\cite{Vollhardt90}.
The pairing symmetry from general point of view has been discussed in the framework of the Ginzburg-Landau (GL) theory\cite{Machida01,Shick01,Mineev02,Dahl07,Shopova09}.
The magnetization process and the structure of the vortex core in the mixed state has been studied\cite{Koyama83,Tachiki79}.

As a direct consequence of the interplay, it has been discussed a possibility of a self-induced vortex state in UCoGe.
Two relaxation frequencies are observed below $T_c$ in $^{59}$Co NQR measurement\cite{Ohta10}, suggesting the contribution from the vortex cores.
The observed magnetization curve indicates a ferromagnetic moment dominates over a diamagnetic contribution due to the supercurrent\cite{Deguchi10}.
These experimental observations provide indirect evidences for the self-induced vortex state in UCoGe.

In this paper, we develop the simple theoretical framework for ferromagnetic superconductors in the (self-induced) vortex state.
For this purpose, we combine the spatially averaged approximation, which has been developed to treat the mixed phase of the type-II superconductors\cite{Brandt67,Pesch75,Houghton98,Kusunose04}, and the phenomenological form of the free-energy functional for the ferromagnetism\cite{Moriya85,Blount79}.
Using the framework, we discuss the thermodynamic properties in the self-induced vortex state.
The obtained expressions of the critical fields are used to determine the occurrence condition for the self-induced vortex state.
The comparison of the calculated quantities with the observed ones indicates that the self-induced vortex state indeed occurs in UCoGe, and its pairing symmetry seems to be the A$_2$-type with a small spin difference in the gap magnitudes.
In this paper, we restrict our attention to the Ising FM + SC state, and single domain of the FM.
The domain structure and the vortex dynamics have been investigated in literatures\cite{Dao11,Lin12}.

The organization of the paper is as follows.
In the next section, we construct the effective free-energy functional of the ferromagnetic superconductors by introducing the spatially averaged approximation.
In \S3, we discuss the critical fields within the approximation, and the occurrence condition for the self-induced vortex state.
The temperature dependence of the specific heat, the magnetization process, and the density of states (DOS) are demonstrated.
We summarize the paper in the last section.
For comparison with the results of the self-induced vortex state as discussed in the main text, the simple mean-field analysis of the coexisting phases without the orbital de-pairing is given in Appendix.

\section{Formulation}

\subsection{Effective free-energy functional for ferromagnetic superconductors}

Let us begin with the quasi-classical Gibbs free-energy functional per unit volume of the gap function $\Delta_{\alpha\beta}$, the vector potential $\mib{A}$, and the magnetization $\mib{M}$ in the form,
\begin{equation}
{\cal F}[\Delta,\mib{A},\mib{M}]={\cal F}_{\rm SC}+{\cal F}_{\rm FM},
\end{equation}
where ${\cal F}_{\rm SC}$ and ${\cal F}_{\rm FM}$ represent the superconducting part including the coupling to the ferromagnetism, and the ferromagnetic part with the field energy, respectively.
Throughout this paper, we restrict our attention to a triplet ESP state, i.e., $\Delta_\sigma\equiv\Delta_{\sigma\sigma}$.
The stability of this state within the mean-field calculation in the Meissner state is briefly discussed in the Appendix.
The ESP state is described by the spin-dependent gap function $\Delta_\sigma(\hat{\mib{k}};\mib{R})$, where the spin quantization axis is taken as being parallel to the magnetic field along $z$ axis $\mib{B}=\mib{\nabla}\times\mib{A}=B\mib{e}_z$ and $\mib{M}=M\mib{e}_z$.
$\Delta_\sigma$, $B$ and $M$ depend on the center-of-mass coordinate $\mib{R}$ of the Cooper pair due to the presence of the (self-induced) vortex lattice.

The pure superconducting state is affected by the FM through (i) the Zeeman coupling (the Pauli de-pairing), (ii) the gauge-invariant coupling (the orbital de-pairing), and (iii) the spin-dependent DOS at the Fermi energy $\rho_{0\sigma}(M)$.
Since we consider the case that the so-called $\mib{d}$-vector is perpendicular to $\mib{B}$, we can neglect the paramagnetic effect of (i), and the effect of (iii) is more important (see also Appendix).
It is discussed that the paramagnetic effect is negligible even for $\mib{H}\perp\mib{e}_z$\cite{Mineev10}.
Then, the superconducting part ${\cal F}_{\rm SC}$ is independent of $M$ except of $\rho_{0\sigma}(M)$, and the effect of the FM arises only through $\mib{A}$ (the effect of (ii)).

With these assumptions, ${\cal F}_{\rm SC}$ for the ESP state in the quasi-classical framework\cite{Eilenberger68,Kusunose04} is given by
\begin{equation}
{\cal F}_{\rm SC}=\frac{1}{2}\sum_\sigma\rho_{0\sigma}\int d\mib{R}\left[
\left\{\ln\frac{T}{T_{c\sigma}}+\sum_{n\ge0}\frac{2\pi T}{\omega_n}\right\}\left\langle\left|\Delta_\sigma\right|^2\right\rangle-\langle I_\sigma \rangle
\right],
\label{fsf1}
\end{equation}
with
\begin{equation}
I_\sigma=-2\pi T\sum_{n\ge0}{\rm Re}\left[
\Delta_\sigma f_\sigma^*+\Delta_\sigma f_\sigma+\frac{f_\sigma^*{\cal L}_+f_\sigma+f_\sigma{\cal L}_-f_\sigma^*}{1+g_\sigma}
\right],
\label{fsf2}
\end{equation}
where ${\cal L}_{\pm}=\omega_n\pm(\mib{v}_{\rm F}/2)\cdot(\mib{\nabla}\mp 2ie\mib{A})$ with the fermionic Matsubara frequency $\omega_n=\pi T(2n+1)$ and the Fermi velocity $\mib{v}_{\rm F}$.
The spatial integral is taken over the unit volume.
$T_{c\sigma}=(2\omega_c e^\gamma/\pi)\exp(-1/\rho_{0\sigma}g)$ for the weak-coupling attraction $g>0$.
$g_\sigma(\hat{\mib{k}},i\omega_n;\mib{R})$ and $f_\sigma(\hat{\mib{k}},i\omega_n;\mib{R})$ are the normal and the anomalous components of the quasi-classical propagator, respectively.
The bracket $\langle\cdots\rangle$ represents the $\hat{\mib{k}}$ average over the Fermi surface.
In this paper, we assume the isotropic Fermi surface for simplicity.

For the ferromagnetic part, we adopt the phenomenological expression for a given external uniform magnetic field $H$ as\cite{Landau-Lifshitz}
\begin{multline}
{\cal F}_{\rm FM}=\int d\mib{R}\,\biggl[\frac{\chi^{-1}}{2}(M-M_0)^2+c|\mib{\nabla}M|^2
\\
+\frac{(B-H-4\pi M)^2}{8\pi}-MH\biggr]-F_{\rm N},
\label{ffm}
\end{multline}
where $M_0$ is the spontaneous uniform magnetization in the normal phase, $\chi^{-1}>0$ is the inverse uniform susceptibility, $c>0$, and ${\cal F}_{\rm FM}$ is measured from that of the normal ferromagnetic state $F_{\rm N}$.
Note that in the normal state the equilibrium values are given by the conditions, $\partial{\cal F}_{\rm FM}/\partial B=0$ and $\partial{\cal F}_{\rm FM}/\partial M=0$, as
\begin{gather}
B=H+4\pi M,
\cr
M=M_0+\chi H.
\label{normb}
\end{gather}
$F_{\rm N}$ is then given by $F_{\rm N}=-(M_0H+\chi H^2/2)$.

Equations~(\ref{fsf1}), (\ref{fsf2}) and (\ref{ffm}) constitute the effective free-energy functional to describe the ferromagnetic superconductors.

\subsection{Spatially averaged approximation}

In order to examine the thermodynamic properties of the ferromagnetic superconductors, we introduce the spatially averaged approximation.
This approximation was originally proposed by Brandt-Tewordt-Pesch\cite{Brandt67} and Pesch\cite{Pesch75} for the superconducting state near the upper critical field $B\lesssim H_{c2}$.
In this approximation, the $\mib{R}$ dependences of $|\Delta_\sigma(\hat{\mib{k}};\mib{R})|^2$, $g_\sigma(\hat{\mib{k}},i\omega_n;\mib{R})$ and $B(\mib{R})$ are approximated by their spatial averages, $\overline{\Delta}_\sigma^2|\varphi_\sigma(\hat{\mib{k}})|^2$, $\overline{g}_\sigma(\hat{\mib{k}},i\omega_n)$ and $\overline{B}$, where $\varphi_\sigma(\hat{\mib{k}})$ is the angular part of the gap function (The normalization is $\langle|\varphi_\sigma(\hat{\mib{k}})|^2\rangle=1$).
The important phase winding in $\Delta_\sigma(\hat{\mib{k}};\mib{R})$ and $f_\sigma(\hat{\mib{k}},i\omega_n;\mib{R})$ due to the vortex cores is taken into account by using the formal expression of the Abrikosov lattice solution.
With these approximations, the quasi-classical Eilenberger equation can be solved, and the analytic expressions of $\overline{g}_\sigma$ and ${\cal F}_{\rm SC}$ can be obtained.
The averaged values $\overline{\Delta}_\sigma$ and $\overline{B}$ are then determined by minimizing the resultant free-energy functional\cite{Kusunose04}.
For the ferromagnetic superconductors, $M(\mib{R})$ is also approximated by its average $\overline{M}$, which is regarded as an additional variational parameter.
Hereafter, we omit the overlines of the spatial averages for notational simplicity.

The analytic expressions are given by
\begin{align}
&{\cal F}_{\rm SC}[\Delta_\sigma,B]=\frac{1}{2}\sum_\sigma \rho_{0\sigma}\Delta_\sigma^2\Biggl[
\ln\frac{T}{T_{c\sigma}}
\cr&\quad\quad\quad\quad\quad\quad
+2\pi T\sum_{n\ge0}\Biggl\langle
\frac{|\varphi_\sigma|^2}{\omega_n}\Biggl(1-\frac{2g_\sigma}{1+g_\sigma}S(\zeta)\Biggr)
\Biggr\rangle
\Biggr],
\\
&\qquad
g_\sigma(\hat{\mib{k}},i\omega_n)=\frac{\omega_n}{\sqrt{\omega_n^2+\Delta_\sigma^2|\varphi_\sigma|^2S_1(\zeta)}},\,\,\,(\omega_n>0),
\\
&{\cal F}_{\rm FM}[B,M]=\frac{\chi^{-1}}{2}(M-M_0)^2+\frac{(B-H-4\pi M)^2}{8\pi}
\cr&\quad\quad\quad\quad\quad\quad\quad\quad\quad\quad\quad\quad\quad\quad\quad\quad
-MH-F_{\rm N},
\end{align}
where $\zeta=v_{\rm F}\sin\theta\,\sqrt{2|e|B}/2i\omega_n$ ($\theta$ is the polar angle from $B$ direction).
Note that the $B$ dependence in ${\cal F}_{\rm SC}$ arises only through $\zeta$.
Here, we have introduced the functions,
\begin{equation}
S(\zeta)=\frac{-i\sqrt{\pi}}{\zeta}W(1/\zeta),
\quad
S_1(\zeta)=\frac{i\sqrt{\pi}}{\zeta^2}W'(1/\zeta),
\end{equation}
in which $W(z)=e^{-z^2}{\rm erfc}(-iz)$ and $W'(z)$ are the Faddeeva function and its derivative.
For $|\zeta|\ll1$ (corresponding to the limits, $B\to0$, $|\omega_n|\to\infty$, $\theta\to0,\pi$), the expansion of $S$ and $S_1$ are given by
\begin{align}
&S(\zeta)=1+\frac{1}{2}\zeta^2+\frac{3}{4}\zeta^4+\frac{15}{8}\zeta^6+\cdots,
\\
&S_1(\zeta)=1+\frac{3}{2}\zeta^2+\frac{15}{4}\zeta^4+\cdots.
\end{align}

By using these expansions, the above analytic expressions reproduce the Meissner limit ($B\to0$) as
\begin{align}
&{\cal F}_{\rm SC}[\Delta_\sigma,0]=\frac{1}{2}\sum_\sigma \rho_{0\sigma}\Biggl[
\Delta_\sigma^2\ln\frac{T}{T_{c\sigma}}
\cr&\quad\quad\quad\quad
+2\pi T\sum_{n\ge0}\omega_n\Biggl\langle
\biggl(\sqrt{1+\Delta_\sigma^2|\varphi_\sigma|^2/\omega_n^2}-1\biggr)^2
\Biggl\rangle
\Biggr],
\\
&\qquad g_\sigma(\hat{\mib{k}},i\omega_n)=\frac{\omega_n}{\sqrt{\omega_n^2+\Delta_\sigma^2|\varphi_\sigma|^2}},\,\,\,(\omega_n>0, B=0).
\end{align}
Although the spatially averaged Pesch approximation was originally intended to describe the mixed phase in the vicinity of $H_{c2}$, it turned out that it recovers the Meissner limit as well.
In this sense, this approximation gives us a practical interpolation scheme, if we appropriately minimize the free-energy functional.

\subsection{Units and dimensionless expressions}

For practical computation, we first introduce the characteristic quantities with $\rho_{0\sigma}=\rho_0$.
The critical temperature $T_c$ at $B=0$ and the upper critical field $B_{c2}$ at $T=0$ within the Pesch approximation\cite{Kusunose04} are given by
\begin{align}
&T_c=\frac{2\omega_ce^\gamma}{\pi}e^{-1/\lambda},
\quad
\lambda=\rho_0g>0,
\quad
(B=0),
\\
&B_{c2}=\frac{2T_c^2}{|e|v_{\rm F}^2}\pi^2\exp[-\langle |\varphi_\sigma|^2\ln \sin^2\theta\rangle-\gamma],
\quad
(T=0).
\end{align}
Then, $\ln(T/T_{c\sigma})=\ln(T/T_c)+(1-\rho_0/\rho_{0\sigma})/\lambda$.
The ratio of the gap magnitude $\Delta_0$ at $T=B=0$ and $T_c$ is given by
\begin{equation}
\frac{\Delta_0}{T_c}=\pi\exp\left[-\frac{1}{2}\langle |\varphi_\sigma|^2\ln|\varphi_\sigma|^2\rangle-\gamma\right].
\end{equation}
In terms of these quantities, the Ginzburg-Landau (GL) parameter is expressed as
\begin{equation}
\kappa^2=\frac{B_{c2}^2}{8\pi\rho_0\Delta_0^2}.
\end{equation}
Within the Pesch approximation, the type-I and the type-II boundary is given by $\kappa_c=1/\sqrt{2}$, which is the same as the GL theory. 

The free-energy functional is measured in unit of $\rho_0\Delta_0^2$, i.e.,
\begin{align}
&\frac{{\cal F}_{\rm SC}}{\rho_0\Delta_0^2}=\frac{1}{2}\sum_\sigma \frac{\rho_{0\sigma}\Delta_\sigma^2}{\rho_{0}\Delta_0^2}\Biggl[
\ln\frac{T}{T_{c\sigma}}
\cr&\quad\quad\quad\quad\quad\quad
+2\pi T\sum_{n\ge0}\Biggl\langle
\frac{|\varphi_\sigma|^2}{\omega_n}\Biggl(1-\frac{2g_\sigma}{1+g_\sigma}S(\zeta)\Biggr)
\Biggr\rangle
\Biggr],
\\
&\frac{{\cal F}_{\rm FM}}{\rho_0\Delta_0^2}=\kappa^2\biggl[u(m-m_0)^2+(b-h-m)^2
-2mh\biggr]-\frac{F_{\rm N}}{\rho_0\Delta_0^2},
\cr
\end{align}
where we have introduced the dimensionless magnetizations and the magnetic fields as $m=4\pi M/B_{c2}$, $m_0=4\pi M_0/B_{c2}$, $h=H/B_{c2}$, $b=B/B_{c2}$.
$u=1/4\pi\chi$ represents the ``distance" from the ferromagnetic quantum critical point (QCP).

Fay and Appel discussed that the superconducting $T_c$ is extremely suppressed near the FM QCP\cite{Fay80}.
The argument was based on the mean-field theory associated with the renormalized characteristic energy of the spin fluctuations confirmed by the analysis of the Eliashberg equation.
However, it has been recognized that the vertex corrections beyond the Eliashberg theory are important near the QCP.
The renormalization-group argument showed that $T_c$ becomes finite at the QCP, for instance\cite{Roussev01}.
Moreover, it is proposed that the FM transition becomes first order near the FM QCP\cite{Chubukov03}.
Note that how close we can approach to the FM QCP in reality is currently under debate.

\subsection{Effect of magnetic polarization}

Now, let us assume the linear $M$ dependence of the DOS, i.e.,
\begin{equation}
\rho_{0\sigma}(M)=\rho_0[1+\alpha(M_0+\chi H)\sigma],
\label{dosm}
\end{equation}
with the constant parameter $\alpha$, in which we have used the value in the normal state (\ref{normb}) for $M(H)$.
Note that $\rho_{0\sigma}$ no longer depends on $M$ by this simplification.

Since ${\cal F}_{\rm SC}$ does not depend on $M$ in our approximation, we can easily obtain the equilibrium value of $M$ by $\partial{\cal F}/\partial M=\partial{\cal F}_{\rm FM}/\partial M=0$.
Namely, we obtain
\begin{equation}
M=\frac{M_0+\chi B}{1+4\pi\chi},
\quad\left(m=\frac{u\,m_0+b}{1+u}\right).
\label{mpol}
\end{equation}
If we combine this result with the Maxwell equation, $4\pi\mib{j}=c\mib{\nabla}\times(\mib{B}-4\pi\mib{M})$, where $\mib{j}$ is the supercurrent, we obtain the effective London penetration depth\cite{Sonin98} as
\begin{equation}
\lambda_{\rm eff}^2=\frac{\lambda^2}{1+4\pi\chi},
\quad
\left(
\lambda_{\rm eff}^2=\frac{u}{1+u}\lambda^2
\right),
\label{relam}
\end{equation}
where $\lambda$ is the penetration depth without magnetic substances, i.e. $\chi=0$.
This indicates that the penetration depth becomes shorter as the system approaches to the QCP ($u\to0$).

Substituting (\ref{mpol}) into the free-energy functional, we obtain
\begin{equation}
\frac{{\cal F}[\Delta_\sigma,B]}{\rho_0\Delta_0^2}=\frac{{\cal F}_{\rm SC}[\Delta_\sigma,B]}{\rho_0\Delta_0^2}+\kappa_{\rm eff}^2\frac{(B-H_{\rm eff})^2}{B_{c2}^2}-\frac{F_{\rm N}}{\rho_0\Delta_0^2},
\label{fdb}
\end{equation}
where we have introduced the effective parameters as
\begin{align}
&H_{\rm eff}=H+4\pi(M_0+\chi H),
\quad
\left(
h_{\rm eff}=m_0+\frac{1+u}{u}h
\right),
\label{heff}
\\
&\kappa^2_{\rm eff}=\frac{\kappa^2}{1+4\pi\chi},
\quad
\left(
\kappa_{\rm eff}^2=\frac{u}{1+u}\kappa^2
\right).
\end{align}
These results indicate that our system of ferromagnetic superconductor is equivalent to that with the GL parameter $\kappa_{\rm eff}$ at the effective ``external" magnetic field $H_{\rm eff}$.
As mentioned in the above, the renormalization factors in $H_{\rm eff}$ and $\kappa_{\rm eff}$ arise from (\ref{normb}) and (\ref{relam}).

After we take into account the effect of the magnetic polarization,
we will minimize the functional (\ref{fdb}) to determine the equilibrium values of $\Delta_\sigma$ and $B$ for given $T$ and $H$ ($H_{\rm eff}$ via (\ref{heff})).
Note that in the London limit ($\kappa\gg1$), $B\to H_{\rm eff}$, and only the two variational parameters $\Delta_\sigma$ are left to be determined.

Once we find the minimum solutions, we can calculate the thermodynamic quantities such as the entropy,
\begin{multline}
\frac{S}{\gamma_{\rm N}T_c}=-\frac{3}{\pi^2}\sum_\sigma\frac{\rho_{0\sigma}}{\rho_0}\int_0^\infty d\omega\,\rho_{\sigma}(\omega)
\times
\\
\times [f(\omega)\ln f(\omega)+(1-f(\omega))\ln(1-f(\omega))],
\end{multline}
and the specific heat is obtained by the numerical differentiation.
Here, $\gamma_{\rm N}=2\pi^2\rho_0/3$ is the Sommerfeld coefficient and $f(x)=1/(e^{x/T}+1)$ is the Fermi-Dirac distribution function.
Here, the spin-dependent DOS in the superconducting state is given by
\begin{equation}
\rho_{\sigma}(\omega)=\rho_{0\sigma}\,{\rm Re}\biggl[\biggl\langle g_\sigma(\hat{\mib{k}},\omega+i0)\biggr\rangle\biggr].
\end{equation}

\section{Results}

\subsection{Critical fields}
Let us first examine the upper, the lower, and the thermodynamic critical fields within the spatially averaged framework for the case without the DOS splitting ($\alpha=0$).

The upper critical field at $T=0$ is determined by the condition $H_{\rm eff}=B_{c2}$, namely,
\begin{equation}
\frac{H_{c2}}{B_{c2}}=\frac{u}{1+u}(1-m_0).
\end{equation}
For $m_0>1$ the system becomes a normal state.
At first glance, $H_{c2}$ seems to be a monotonic increasing function of $u$.
In practice, however, the spontaneous magnetization $m_0$ itself depends on $u$, and it is usually an increasing function as well.
For instance, in the simple mean-field theory\cite{Moriya85}, $m_0$ is proportional to $u^{1/2}$.
Thus, $H_{c2}$ curve as a function of $u$ has a maximum at certain $u$ due to the factor $1-m_0(u)$.

In the Pesch approximation, the free energy $F[B]$ for small $B$, which is obtained by putting the equilibrium values of $\Delta_\sigma$ into (\ref{fdb}) for a fixed $B$, has the form,
\begin{equation}
\frac{F[B]}{\rho\Delta_0^2}=\kappa_{\rm eff}^2\frac{(B-H_{\rm eff})^2}{B_{c2}^2}+f_0+f_1\frac{B}{B_{c2}}+f_2\frac{B^2}{B_{c2}^2}+\cdots,
\end{equation}
where the coefficients $f_n$ are $O(1)$ quantities.
Then, the lower critical field is determined by the sign change of the slope at $B=0$, i.e., $(\partial F/\partial B)\bigr|_{B=0,H=H_{c1}}=0$, namely,
\begin{equation}
\frac{H_{c1}}{B_{c2}}=\frac{f_1}{2\kappa^2}-\frac{u}{1+u}m_0.
\end{equation}

The thermodynamic critical field is given by the condition, $F(B=0)=-F_{\rm N}$ at $H_c$.
Noting the free energy $F_{\rm SC}(B=T=0)=-\rho_0\Delta_0^2/2$, we obtain
\begin{equation}
\frac{H_c}{B_{c2}}=\frac{1}{\sqrt{2}\kappa}\sqrt{\frac{u}{1+u}}-\frac{u}{1+u}m_0.
\end{equation}
Note that in order to satisfy the relation $H_{c1}=H_{c2}=H_c$ for $\kappa=\kappa_c$ and $u\to\infty$ ($\chi\to0$), we identify $f_1=1$.
In reality, $f_1$ slightly deviates from unity due to the artifact of the present approximation.

\subsection{Condition for self-induced vortex state}

\begin{figure}[t]
\begin{center}
\includegraphics[width=7cm]{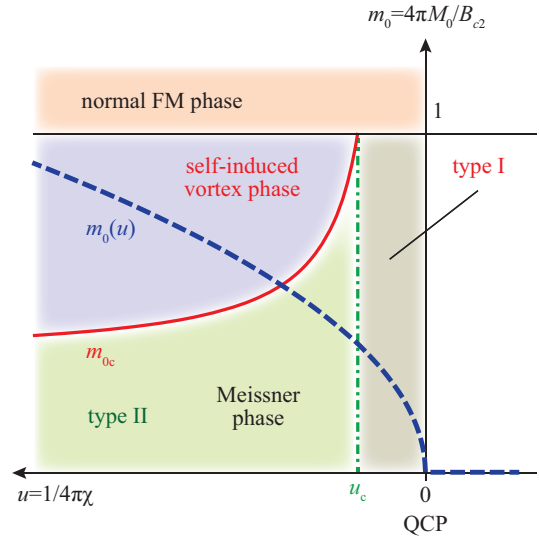}
\end{center}
\caption{(Color online) The possible FM-SC states in the $m_0$-$u$ plane. The self-induced vortex state appears in the region, $m_{0c}<m_0<1$. In the region $u<u_c$, the superconductor exhibits the type-I behavior. The typical $u$ dependence of the spontaneous magnetization $m_0$ is indicated schematically by the dashed line.}
\label{phase}
\end{figure}

Since the effective GL parameter $\kappa_{\rm eff}$ becomes small as $u\to0$, the ferromagnetic superconductor could exhibit the transition from the type II ($u>u_c$) behavior to the type I ($u<u_c$) behavior.
The critical value of $u$ is given by
\begin{equation}
u_c=\frac{1}{2\kappa^2-1}.
\end{equation}

The self-induced vortex state can appear only for $u>u_c$, and the lower critical field satisfies $H_{c1}<0$.
The latter condition gives the critical magnetization $m_{0c}$ for the self-induced vortex state ($m_0>m_{0c}$),
\begin{equation}
m_{0c}=\frac{f_1}{2\kappa^2}\left(1+\frac{1}{u}\right).
\end{equation}

The above results are summarized in the $m_0$-$u$ plane as shown in Fig.~\ref{phase}, in which we have chosen $f_1=1$.
The schematic $u$ dependence of $m_0$ (e.g., $\propto u^{1/2}$) is represented by the dashed line.
As the system at $H=0$ approaches to the QCP along the dashed line, it changes subsequently from the normal ferromagnetic state, the self-induced vortex state, the type-II Meissner state, and then the type-I Meissner state.
In practice, however, a tuning of the interaction for the FM and the SC, e.g., by applying the pressure, may change the electronic structure itself and consequently relevant parameters.
It would result in the complicated $u$ dependence of $m_0$.
In this sense, $u$ and $m_0$ can be regarded as the independent parameters.

\begin{figure}[t]
\begin{center}
\includegraphics[width=8.5cm]{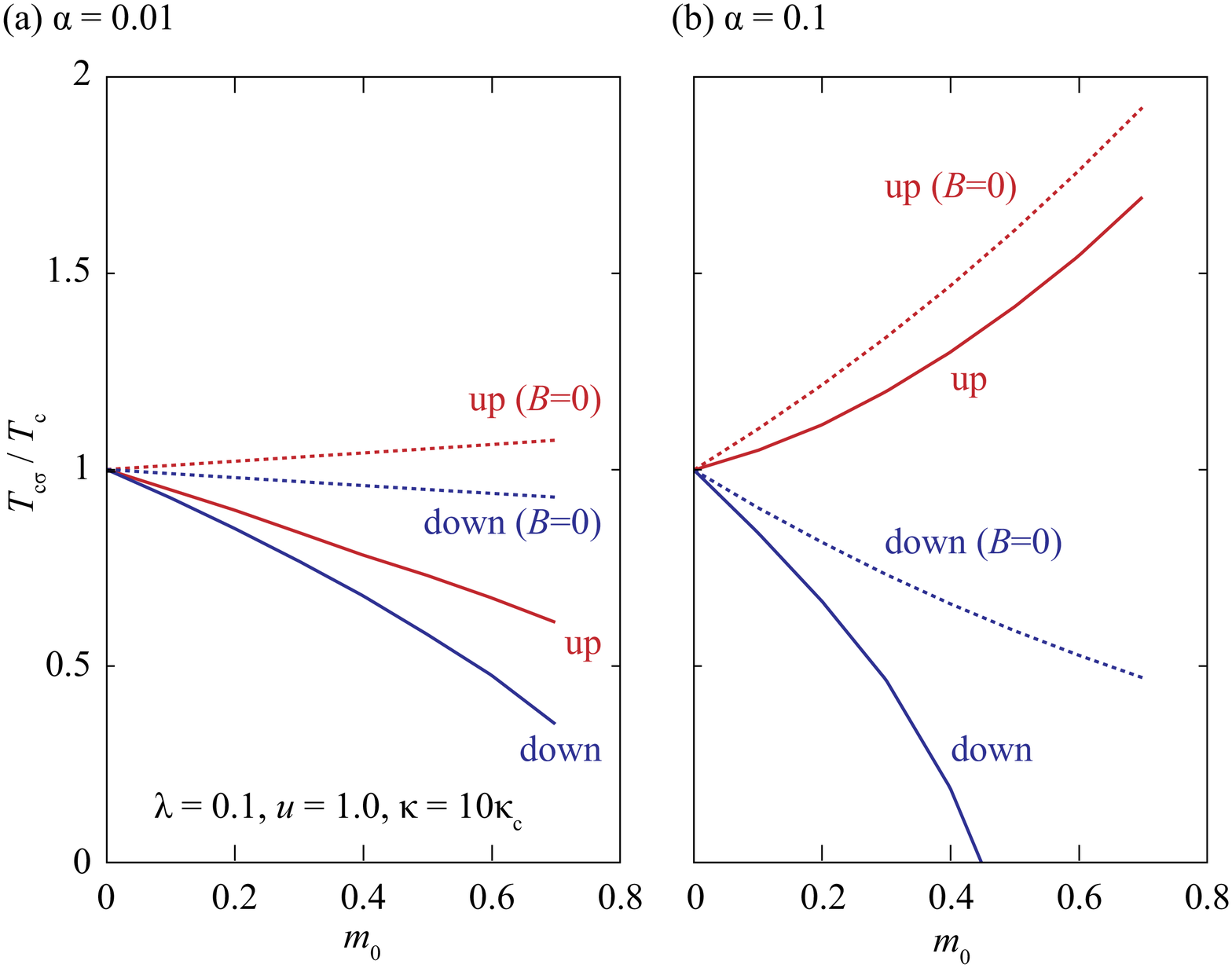}
\end{center}
\caption{(Color online) The $m_0$ dependence of $T_{c\sigma}$ ($\sigma=\uparrow$, $\downarrow$) for (a) the small DOS splitting $\alpha=0.01$, and (b) the large DOS splitting $\alpha=0.1$.}
\label{tc}
\end{figure}

\subsection{Transition temperature}

Hereafter we use the explicit model to investigate the thermodynamic properties of the ferromagnetic superconductor.
We consider the $p$-wave ESP state by using the angular dependence, $\varphi_\sigma=\sqrt{3/2}\sin\theta\, e^{i\sigma\phi}$, and we fix the parameters, $\lambda=0.1$, $u=1.0$, and $\kappa=10\kappa_c$.
In this case, we obtain
\begin{align}
&\frac{\Delta_0}{T_c}=\pi e^{-(3\ln6-5)/6-\gamma}\simeq 1.65693,
\\
&\frac{B_{c2}}{2T_c^2/|e|v_{\rm F}^2}=\frac{\pi^2}{4}e^{5/3-\gamma}\simeq 7.3347.
\end{align}
Since the ESP state is a kind of the two-gap superconductor without the inter-band pair scattering, there appears a double transition corresponding to the larger DOS of the up spin and the smaller DOS of the down spin as (\ref{dosm}) with $\alpha>0$.
Figure~\ref{tc} shows $T_c$ according to the BCS formula for the Meissner state (the dotted lines, $B=0$), and the calculated $T_c$ in the self-induced vortex state (the solid lines) for (a) the small DOS splitting, $\alpha=0.01$, and (b) the large DOS splitting, $\alpha=0.1$.
As $T$ decreases at fixed $m_0$, the FM + SC with A$_1$ state ($\Delta_\uparrow\ne0$, $\Delta_\downarrow=0$) first appears above $T_{c\downarrow}$ and then the FM + SC with A$_2$ state ($\Delta_\uparrow>\Delta_\downarrow\ne0$) appears.
For the large DOS splitting as shown in Fig.~\ref{tc}(b), only the A$_1$ state appears for $m_0\gtrsim 0.45$ over the whole temperature range.

\begin{figure}[t]
\begin{center}
\includegraphics[width=8.5cm]{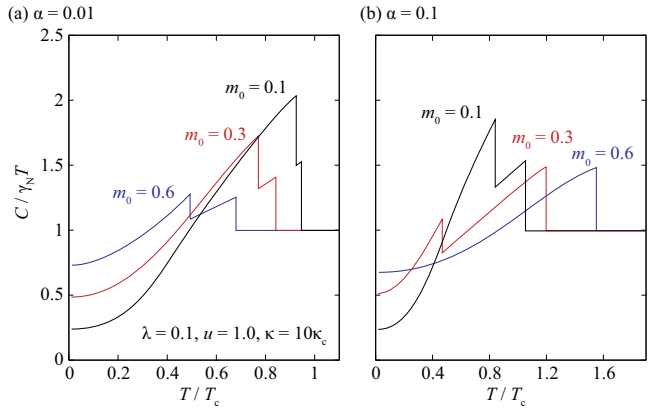}
\end{center}
\caption{(Color online) The $T$ dependence of the specific heat for (a) the small DOS splitting $\alpha=0.01$, and (b) the large DOS splitting $\alpha=0.1$.}
\label{ct}
\end{figure}

\subsection{Specific heat}

\begin{figure}[t]
\begin{center}
\includegraphics[width=8cm]{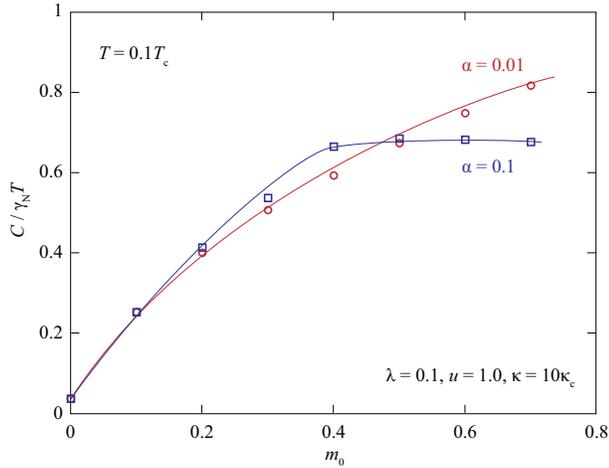}
\end{center}
\caption{(Color online) The calculated residual $\gamma$ values (at $T=0.1T_c$) as a function of the spontaneous magnetization $m_0$. The solid lines are a guide for the eye.}
\label{g0m0}
\end{figure}

\begin{figure}[t]
\begin{center}
\includegraphics[width=8.5cm]{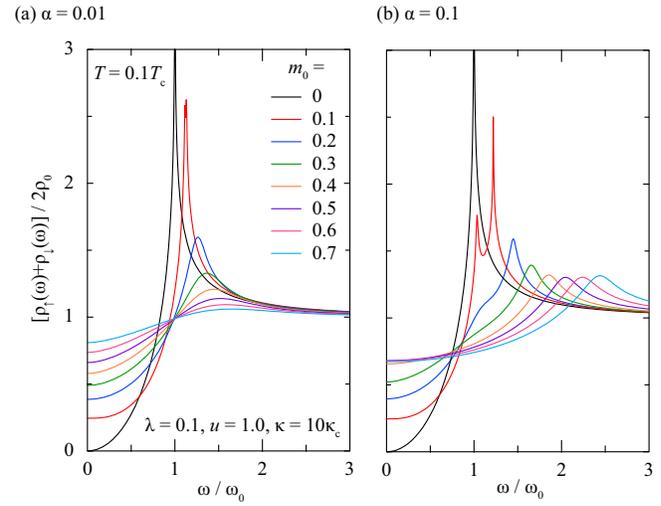}
\end{center}
\caption{(Color online) The total DOS with various $m_0$ at $T=0.1T_c$ for (a) the small DOS splitting $\alpha=0.01$, and (b) the large DOS splitting $\alpha=0.1$. We have introduced $\omega_0=\sqrt{3/2}\Delta_0$.}
\label{dosm0}
\end{figure}

The $T$ dependence of the specific heat is shown in Fig.~\ref{ct} both for the large and the small DOS splittings.
As $m_0$ increases, the double transition is separated with each other, and the residual $\gamma$ coefficient increases due to the contribution from the vortex cores.
The $m_0$ dependence of the residual $\gamma$ value is shown in Fig.~\ref{g0m0}, which is qualitatively similar to the observed tendency\cite{Aoki13}.
Note that if the origin of the residual $\gamma$ value is the fully polarized gap, i.e., the A$_1$ state, the residual $\gamma$ value should always be larger than $50\%$, and simultaneously the trace of the double transition should be observed.

Through the comparison among three typical ferromagnetic superconductors, UGe$_2$, URhGe, and UCoGe, the larger $m_0$ yields the larger residual $\gamma$ value\cite{Aoki13}.
Additionally, these superconductors exhibit relatively broad transitions in the specific heat rather than double transition.
The observed residual $\gamma$ value is about $15\%$ in UCoGe\cite{Aoki13}.
These facts are consistent with our results accompanied by the self-induced vortex lattice.
We eventually expect the small DOS splitting due to the ordered moment in UCoGe.

The $m_0$ dependence of the total DOS in the ferromagnetic superconducting state is shown in Fig.~\ref{dosm0}.
For the larger DOS splitting, there is a distinct two-peak structure for small $m_0$.
The zero-energy DOS for $\alpha=0.1$ shows a saturating tendency for large $m_0$, which is reflected in the saturating behavior of the residual $\gamma$ values in Fig.~\ref{g0m0} (the open squares).

\subsection{Magnetization curve}

\begin{figure}[t]
\begin{center}
\includegraphics[width=8.5cm]{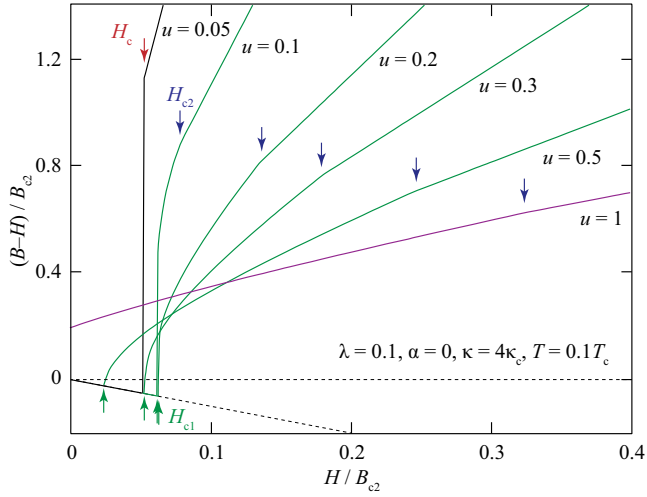}
\end{center}
\caption{(Color online) The magnetization curve for various $u$ and $m_0(u)=0.3\,u^{1/2}$. The arrow indicates the critical fields. The case $u=0.05$ shows the type-I behavior, while the cases $u\ge 0.1$ exhibit the type-II behavior. The typical behavior in the self-induced vortex state is shown for $u=1$.}
\label{mh}
\end{figure}

Finally, we discuss the magnetization curve in the ferromagnetic superconductor.
Figure~\ref{mh} shows the magnetization ($B-H$) curve for $\kappa=4\kappa_c$, in which both the ordered moment and the diamagnetic supercurrent contribute to $B-H$.
Here, we have used an example mean-field $u$ dependence of $m_0$, i.e., $m_0(u)=0.3u^{1/2}$.
As expected in Fig.~\ref{phase}, the magnetization curves show the type-I behavior for $u=0.05$, while for $u=0.1\sim 0.5$, it shows the type-II behavior.
The typical behavior in the self-induced vortex state is obtained for $u=1$, which is indeed observed in the magnetization measurements of UCoGe\cite{Deguchi10}.
If a system is characterized by considerably small $\kappa$, and it is sufficiently close to the FM QCP, we would expect to observe the transition from the type-II to the type-I behavior by applying a pressure.

\section{Summary}

We have developed the quasi-classical and semi-phenomenological theory for ferromagnetic superconductors.
Neglecting the Pauli de-pairing, and using the spatially averaged approximation, we have obtained the effective free-energy functional to determine the thermodynamic properties in the self-induced vortex state.
The effect of the magnetic polarization can be absorbed into the effective values of the Ginzburg-Landau parameter and the external magnetic field.
For systems with large magnetic susceptibility, their penetration depth and consequently their GL parameter become smaller, yielding the transition from the type-II to the type-I superconductor by approaching to the FM QCP.

The detailed calculation for the $p$-wave ESP state have revealed that the observed behaviors in the specific heat and the magnetization curve are compatible with the self-induced A$_2$ vortex state rather than the fully polarized A$_1$ state.
It is also expected that the difference of the gap magnitude between the opposite spins is small in UCoGe.
A direct evidence from the microscopic experimental probes, such as NMR and neutron scattering is highly desired for deeper understanding of this peculiar coexisting state in ferromagnetic superconductors.

\section*{Acknowledgments}
We would like to thank D. Aoki, J. Flouquet, K. Hattori, M. Sigrist and M. Matsumoto for fruitful discussions.
This work was supported by a Grant-in-Aid for Scientific Research on Innovative Areas ``Heavy Electrons" (No.20102008) from The Ministry of Education, Culture, Sports, Science, and Technology (MEXT), Japan, and for Scientific Research C (No. 23540414) from the Japan Society for the Promotion of Science.

\appendix

\section{Mean-Field Analysis of Coexisting Phases}

\begin{figure}[t]
\begin{center}
\includegraphics[width=8.5cm]{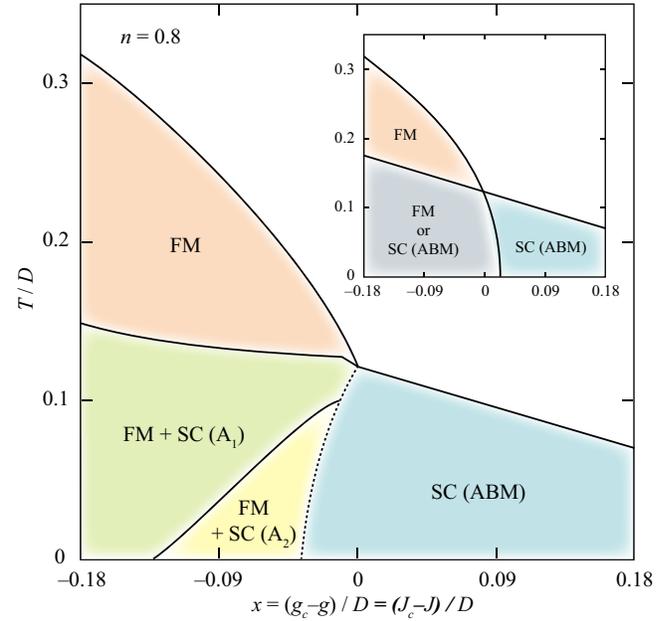}
\end{center}
\caption{(Color online) The mean-field phase diagram for $n=0.8$. The coupling constants $g$ and $J$ are changed simultaneously from $g_c$ and $J_c$, which are chosen such that $T_{\rm FM}$ or $T_c$ of the pure state becomes $0.12D$. The inset shows $T_{\rm FM}$ or $T_c$ without the another phase. The solid  lines indicate the 1st-order phase transition, while the dotted line represents the 2nd-order phase transition.}
\label{mfphase}
\end{figure}
\begin{figure}[t]
\begin{center}
\includegraphics[width=8.5cm]{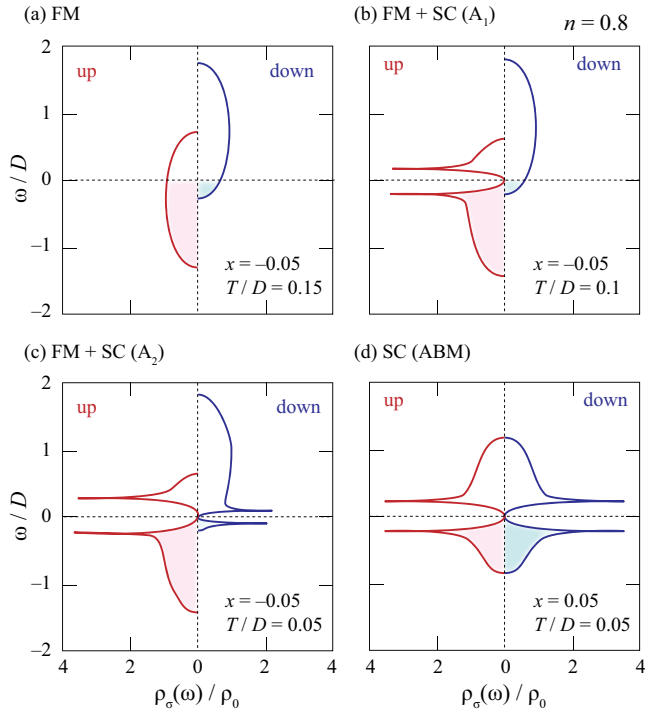}
\end{center}
\caption{(Color online) The spin-dependent DOS for $n=0.8$ in (a) the FM phase, (b) the FM + SC (A$_1$) phase, (c) the FM + SC (A$_2$) phase, (d) the SC (ABM) phase. $x$ and $T/D$ indicate the parameters at which the DOS is calculated.}
\label{mfdos}
\end{figure}

In this Appendix, we perform the simple mean-field analysis\cite{Nevidomskyy05} of the coexisting phases near the FM QCP.
For brevity, we consider the Meissner state and neglect the effect of the orbital de-pairing.

Let us start with the simplest Hamiltonian for the ferromagnetic ESP superconductivity,
\begin{multline}
H=\sum_{\mib{k}\sigma}\xi_k c_{\mib{k}\sigma}^\dagger c_{\mib{k}\sigma}^{}-J\sum_{\mib{q}}s_{\mib{q}}^zs_{-\mib{q}}^z
\\
-\frac{g}{2}\sum_{\mib{k}}\sum_\sigma^\pm \varphi_\sigma(\hat{\mib{k}})\varphi_\sigma^*(\hat{\mib{k}}')
c_{\mib{k}\sigma}^\dagger c_{-\mib{k}\sigma}^\dagger c_{-\mib{k}'\sigma}^{} c_{\mib{k}'\sigma}^{},
\end{multline}
where $s_{\mib q}^z=\sum_\sigma \sigma c_{\mib{k}+\mib{q}\sigma}^\dagger c_{\mib{k}\sigma}^\dagger$, and $J$, $U>0$.

The self-consistent equations for the ferromagnetic moment $M=\langle s_0^z\rangle_{\rm MF}=n_\uparrow-n_\downarrow$ and the superconducting gap $\Delta_\sigma=-g\sum_{\mib{k}'}\varphi_\sigma(\hat{\mib{k}}') \langle c_{-\mib{k}'\sigma}c_{\mib{k}'\sigma}\rangle_{\rm MF}$ are given by
\begin{align}
&n_\sigma=\frac{1}{2}-T\sum_n\sum_{\mib{k}}\frac{\xi_k-JM\sigma}{\omega_n^2+(\xi_k-JM\sigma)^2+\Delta_\sigma^2|\varphi_\sigma|^2},
\\
&\Delta_\sigma=gT\sum_n\sum_{\mib{k}}\frac{\Delta_\sigma|\varphi_\sigma|^2}{\omega_n^2+(\xi_k-JM\sigma)^2+\Delta_\sigma^2|\varphi_\sigma|^2}.
\end{align}
The Landau free energy measured from the paramagnetic normal state is obtained as
\begin{multline}
F=-\frac{T}{2}\sum_{n\mib{k}\sigma}\ln\left(\frac{\omega_n^2+(\xi_k-JM\sigma)^2+\Delta_\sigma^2|\varphi_\sigma|^2}{\omega_n^2+\xi_k^2}\right)
\\
+\frac{1}{2g}\sum_\sigma \Delta_\sigma^2+\frac{J}{2}M^2,
\end{multline}
where $\xi_k=\epsilon_k-\mu$, and the chemical potential $\mu$ is determined by the fixed $n=n_\uparrow+n_\downarrow$.

Expanding $F$ with respect to $M$ and $\Delta_\sigma$, we obtain the Landau expansion as
\begin{multline}
F=a\sum_\sigma \Delta_\sigma^2+b\sum_\sigma\Delta_\sigma^4+a'M^2+b'M^4
\\
+cM\sum_\sigma\sigma\Delta_\sigma^2+c'M^2\sum_\sigma\Delta_\sigma^2+\cdots,
\end{multline}
where
\begin{align}
&a=\frac{1}{2}\left(\frac{1}{g}-T\sum_{n\mib{k}}\frac{1}{\omega_n^2+\xi_k^2}\right),
\quad
b=\frac{T}{4}\sum_{n\mib{k}}\frac{|\varphi_\sigma|^4}{(\omega_n^2+\xi_k^2)^2},
\cr
&a'=J^2\left(\frac{1}{2J}-T\sum_{n\mib{k}}\frac{\omega_n^2-\xi_k^2}{(\omega_n^2+\xi_k^2)^2}\right),
\cr
&b'=\frac{J^4T}{2}\sum_{n\mib{k}}\frac{\omega_n^4-6\omega_n^2\xi_k^2+\xi_k^4}{(\omega_n^2+\xi_k^2)^4},
\cr
&c=-JT\sum_{n\mib{k}}\frac{\xi_k}{(\omega_n^2+\xi_k^2)^2},
\quad
c'=\frac{J^2T}{2}\sum_{n\mib{k}}\frac{\omega_n^2-3\xi_k^2}{(\omega_n^2+\xi_k^2)^3}.
\end{align}
We have assumed that $|\varphi_\sigma|$ is independent of $\sigma$.
It should be noted that the $c$-term causes the difference of the gap magnitude between the opposite spins, which vanishes in the presence of the particle-hole symmetry.
The relative sign between $M$ and $\Delta_\uparrow^2-\Delta_\downarrow^2$ is determined by the sign of $c$\cite{Dahl07}.
As long as the $c$-term is small, e.g., an almost flat DOS, it is less important than the DOS splitting $\rho_{0\sigma}(M)$ due to the Stoner shift,
which results in the exponential difference of $T_c$'s between the opposite spins.
Notice that the quasi-classical theory as that in the main text always assumes the particle-hole symmetry.
Thus, the contribution from the $c$-term is dropped out.

We adopt the semi-circular DOS, $\rho_0(\epsilon)=\rho_0\sqrt{1-(\epsilon/D)^2}$ for $|\epsilon|<D$, $\varphi_\sigma=\sqrt{3/2}\sin\theta e^{i\sigma\phi}$, and the electron density is fixed as $n=0.8$.
Note that for $n=1$ ($\mu=0$) there exist no FM-SC coexisting phases due to the particle-hole symmetry (vanishing the $c$-term in the Landau expansion).
For $n>1$, the results are the same as those of the system with the exchange of the particle-hole and the spin directions, simultaneously.

Figure~\ref{mfphase} shows the overall phase diagram, where the coupling constants $g$ and $J$ are changed simultaneously.
The values $g_c$ and $J_c$ are chosen such that $T_{\rm FM}$ or $T_c$of the pure state becomes $0.12D$ (see the inset).
Below the pure FM state,  the FM + A$_1$ ($\Delta_\uparrow\ne0$, $\Delta_\downarrow=0$) SC state appears, and the latter $T_c$ is slightly pushed down due to the coupling to the ferromagnetic moment.
By a similar reason, the ABM ($\Delta_\uparrow=\Delta_\downarrow\ne0$) SC state pushes out to the FM region.
The competition between the FM and the SC results in the occurrence of the FM + A$_2$ ($\Delta_\uparrow>\Delta_\downarrow\ne0$) SC state in between.
The typical features of the DOS for each phases are shown in Fig.~\ref{mfdos}.

It should be noted that the mean-field approximation for the FM always overestimates the magnitude of the ordered moment, and it is hard to describe the itinerant weak FM as was observed in UCoGe.
If quantum fluctuations are taking into account properly to reduce the ordered moment, e.g. by means of the Self-Consistent Renormalization (SCR) theory\cite{Moriya85}, the FM + A$_1$ SC phase (the FM + A$_2$ SC phase) is expected to shrink (expand) as compared with those obtained by the mean-field calculation.

\end{document}